# All-MgB$_2$ Josephson tunnel junctions

K. Ueda, S. Saito, K. Semba and T. Makimoto

NTT Basic Research Laboratories, NTT Corporation,

3-1 Wakamiya, Morinosato, Atsugi-shi, Kanagawa 243-0198, Japan

M. Naito

Department of Applied Physics, Tokyo University of Agriculture and Technology,

2-24-16 Naka-cho Koganei-city Tokyo, 184-8588, Japan

Sandwich-type all-MgB$_2$ Josephson tunnel junctions (MgB$_2$/AlO$_x$/MgB$_2$) have been fabricated for the first time with as-grown MgB$_2$ films formed by molecular beam epitaxy. The junctions exhibit substantial superconducting current (I$_c$R$_N$ product ~ 0.8 mV at 4.2K), a well-defined superconducting gap ($\Delta$ = ~2.3 mV), and clear Fraunhofer patterns. The superconducting gap voltage of $\Delta$ agrees well with the smaller gap in the multi-gap scenario. The results demonstrate that MgB$_2$ has great promise for superconducting electronics that can be operated at $T$ ~ 20 K.

Integrated circuits made of superconductors are uniquely suitable for processing digital information at ultrahigh speeds and with very little power dissipation.   However, superconducting circuits must be operated at temperatures that are less than about half of the transition temperature, $T_c$.   In the case of Nb-based circuits, the operating temperature should be close to 4.2 K.   This requires liquid He or large-scale cryocoolers, which is unacceptable for most electronic applications. High-temperature superconductors (HTS) *would* have solved this problem, but no method for fabricating HTS Josephson junctions reproducibly with small variations in device parameters has yet been established after almost twenty years since the discovery.   Here, we report the fabrication of Josephson tunnel junctions using the recently discovered 40 K superconductor $MgB_2$ [1].   There have already been several reports on the fabrication of various types of $MgB_2$ junctions, including point-contact or break junctions [2-4], nanobridges [5], planar junctions by localized ion damage in thin films [6] and ramp-type junctions [7].   However, there is no report on the fabrication of sandwich-type all-$MgB_2$ SIS (Superconductor/Insulator/Superconductor) tunnel junctions ($MgB_2$/Insulator/$MgB_2$), although Saito et al. and Carapella et al. reported the fabrication of sandwich-type tunnel junctions comprising one $MgB_2$ electrodes and another superconductors such as Nb or NbN, respectively [8, 9].   Sandwich-type tunnel junctions are very important from the viewpoint of the integration of superconducting circuits.   In this article, we report that Josephson tunnel junctions using as-grown $MgB_2$ films have successfully been fabricated reproducibly for the first time.   The junctions exhibit substantial superconducting current, a well-defined superconducting gap, and clear Fraunhofer patterns.   The results demonstrate that $MgB_2$ has great promise for superconducting electronics that can be operated at $T \sim 20$ K, which is easily accessible by compact cryocoolers.

Josephson tunnel junctions ($MgB_2$/$AlO_x$/$MgB_2$) were fabricated on sapphire -C substrates by a standard all-in-situ process using as-grown $MgB_2$ films formed at the substrate temperature of below 300°C [10, 11].    Four layers, Au/MgB$_2$/AlO$_x$/MgB$_2$, were deposited *in-situ* in a UHV chamber.    Junctions of 25 μm x 25 μm and 100 μm x 100 μm in area were then fabricated by a standard photolithographic process and Ar ion milling.    The MgB$_2$ films were c-axis oriented and the superconducting transition temperatures were 32 ~ 35 K.    The AlO$_x$ barrier was prepared by depositing 1.4-nm Al metal and oxidizing it in a load-lock chamber, which is essentially identical to the standard Nb technology.    The current- voltage (I-V) characteristics were measured using the standard four-probe method down to 45 mK in a dilution refrigerator.

Figure 1 exhibits typical current-voltage (I-V) characteristics and differential conductance curves as a function of bias voltage (dI/dV-V) for two MgB$_2$/AlO$_x$/MgB$_2$ junctions measured at 4.2 K. The dI/dV-V curves were digitally calculated from I-V characteristics.    The junction area was 100 μm x 100 μm (Fig. 1(a): junction A) and 25 μm x 25 μm (Fig. 1(b): junction B), respectively.    The I-V characteristics measured at 45 mK (Fig. 1(a)) and 61 mK (Fig. 1(b)) for these junctions are also shown in Fig. 1.    The junctions showed typical SIS tunneling characteristics.    The superconducting gap was clearly observed at around 4.5 mV (=2$\Delta$), which agrees well with the smaller gap ($\Delta_S$) in the multi-gap scenario of MgB$_2$ [12, 13].    The subgap leakage current is also small, and almost negligible below 2.0 mV (=$\Delta$), especially in junction B.    The R$_N$/R$_{sg}$ (normal resistance/subgap resistance) measured at superconducting gap voltage of 2.0 mV and at 4.2 K is 5.4 for junction A and 21.8 for junction B.    The superconducting current (I$_c$) is observed in both junctions.    The I$_c$ and normal resistance (R$_N$) are 12.5 μA and 61 Ω for junction A, and 0.4 μA and 830 Ω at 4.2 K for junction B.    The I$_c$ measured at 45 mK for junction A and at 61 mK for junction B is 28.0 and 2.0 μA, respectively.    The I$_c$R$_N$ product of these junctions is ~0.8 mV for junction A and ~0.4 mV for junction B at 4.2 K, and it increases to ~1.7 mV at ~50 mK for both junctions. These values are substantial, but somewhat smaller than the calculated results (~ 4 mV) [14].    The reason for the small $I_cR_N$ product is not clear, but deterioration of the superconducting properties of $MgB_2$ at the interface with $AlO_x$ is the most probable reason.

Figure 2 (a) shows the dc magnetic field dependence of $I_c$ in junction A measured at 45 mK. The magnetic field was applied parallel to the surface of the junction. $I_c$ was suppressed as the external magnetic field increased and showed a familiar Fraunhofer pattern. The experimental curve agrees with the calculated one, although there is a small deviation, which may indicate small nonuniformity of $AlO_x$ barrier layer thickness.

The temperature dependence of $I_c$ ($I_c$-T) for the junction A is plotted in Fig. 2 (b). The superconducting current $I_c$ remained finite up to ~20 K. The $I_c$-T did not follow the Ambegaokar-Baratoff theory predicted for ideal SIS junctions, but showed an exponential-like increase with decreasing temperature. This behavior is often observed in junctions with SN (superconductor-normal metal) boundary, indicating the existence of a normal-metal layer between superconducting $MgB_2$ layer and barrier in our junctions [15]. The deterioration of the $MgB_2$ layers at the interface with $AlO_x$ barriers and/or insufficient oxidation of Al metal layers is most probable cause of the behavior. However, Brinkman et al. claimed that the positive curvature should observe in the $I_c$-T curve of $MgB_2$ SIS junctions with tunneling along c-axis direction due to the two-gap superconducting nature of $MgB_2$ [14]. We need higher quality SIS junctions with larger $I_c$ and smaller subgap leakage current to clarify whether these $I_c$-T behaviors are intrinsic to $MgB_2$ SIS junctions or not.

Figure 3(a) shows the differential conductance curves as a function of bias voltage (dI/dV-V) for junction A measured from temperature of 1.4 to 36.1 K. The gap closes between 30.1 and 36.1 K, which is close to the $T_c$ of our as-grown $MgB_2$ films (~33 K). Figure 3(b) shows the temperature dependence of the superconducting gap voltage ($\Delta$) of the junction as determined by the positions of the peaks of dI/dV-V curves. The solid lines in figure 3 show the calculated temperature dependence of the smaller gap ($\Delta_S$) using the parameters ($\Delta$ =2.3 mV, $T_c$= ~32 K) and the formula in reference 13. The experimental and calculated results agree well. Above 10 K, however, there was extra conductance below $\Delta$. We think the origin of these extra structures is the gap smearing which produces an enhanced quasiparticle population at low energy as observed in point-contact results [2].

It has been reported that MgB$_2$ has two gaps, a larger one ($\Delta_L$) of 6-7 meV and a smaller one ($\Delta_S$) of 2-3 meV [12, 13, 16-20]. The band calculations indicate that the Fermi surface of MgB$_2$ consists of two-dimensional (2D) cylinders from the boron $sp^2$ $\sigma$ bands and three-dimensional (3D) sheets from the $\pi$ bands [12, 13]. The $\sigma$ bands have strong electron-phonon interaction and a large 2D superconducting gap ($\Delta_L$). In contrast, the $\pi$ bands have weaker electron-phonon interaction and a small 3D gap ($\Delta_S$). The use of the larger gap is prerequisite in high-frequency applications. For example, in SIS mixers or digital devices, the upper limit ($f_g$) of the response frequency is proportional to the superconducting energy gap. The $f_g$ for existing superconducting devices fabricated using Nb ($\Delta$= 1.5 meV) is ~700 GHz. To achieve the maximum performance of MgB$_2$ devices, we have to extract the larger gap, which would provide $f_g$ of ~3 THz. However, only the smaller gap (2~3 meV: $f_g$ = ~1.4 THz) was observed in our present junctions. The reason may lie in our junction geometry. Our MgB$_2$ films were c-axis oriented, and the supercurrent flows in the c-axis direction. In this geometry, the dominant contribution to the supercurrent flows comes from the weaker 3D component. "Larger-gap" Josephson junctions may be produced by geometry where the supercurrent flows in the ab-plane [21]. This will be our future work.

Finally, we should mention the contrast between high-$T_c$ cuprates and MgB$_2$ in the difficulty in fabricating Josephson tunnel junctions. In high-$T_c$ cuprates, even after the almost twenty years since their discovery, there has been essentially no report of reliable fabrication of any kind of artificial tunnel junctions, either SIS or even SIN (Superconductor/Insulator/Normal-metal). With regard to the tremendous difficulty in fabricating tunnel junctions using high-$T_c$ cuprates, we have pointed out the intrinsic and serious problem of redox reaction at interface [22]. In contrast, significant steps have already been achieved in the fabrication of $MgB_2$-based Josephson junctions, indicating that there seems to be no intrinsic material problem with $MgB_2$. Hence, we believe $MgB_2$ may be promising for superconducting electronics applications in spite of not having $T_c$ as high as high-$T_c$ cuprates. However, we have to keep in mind that many practical problems remain, especially involving the two-gap superconducting nature of this compound.

In summary, we report first results on fabricating sandwich-type Josephson tunnel junctions ($MgB_2/AlO_x/MgB_2$) using as-grown $MgB_2$ thin films formed by molecular beam epitaxy. The junctions exhibit substantial superconducting current, a well-defined superconducting gap, and clear Fraunhofer patterns. These results suggest $MgB_2$ have a high potential for superconducting electronics that can be operated at $T \sim 20$ K.

The authors thank Drs. H. Sato, H. Yamamoto, S. Karimoto, H. Shibata, M. Ueda and J. Nitta for fruitful discussions, and Drs. K. Torimitsu and H. Takayanagi for their support and encouragement throughout the course of this study.

Figure captions

Fig. 1:   Current-voltage characteristics and differential conductance curves as a function of voltage for $MgB_2/AlO_x/MgB_2$ junctions with an area of (a) 100 μm x 100 μm and (b) 25 μm x 25 μm measured at 4.2 K.   The current-voltage characteristics measured at 45 mK (Fig. 1(a)) and 61 mK (Fig. 1(b)) of the junctions are also included by dotted lines.

Fig. 2: (a) Experimental and calculated dc magnetic field dependence of superconducting critical current ($I_c$) in the $MgB_2/AlO_x/MgB_2$ junction (area: 100μm x 100μm) measured at 45 mK.   Here, $\Phi_0$ is a flux quantum, $\Phi_0 = hc/2e$.   (b) Temperature dependence of maximum $I_c$ for $MgB_2/AlO_x/MgB_2$ junctions with an area of 100 μm x 100 μm.

Fig. 3: (a) Differential conductance curves as a function of voltage (dI/dV-V) for the $MgB_2/AlO_x/MgB_2$ junction (area: 100μm x 100μm) measured from 1.4 to 36.1 K.   (b) The temperature dependence of superconducting gap voltage ($\Delta$) for the junction.   The solid line shows the calculated temperature dependence of $\Delta$.

K. Ueda, S. Saito, K. Semba, T. Makimoto, M. Naito

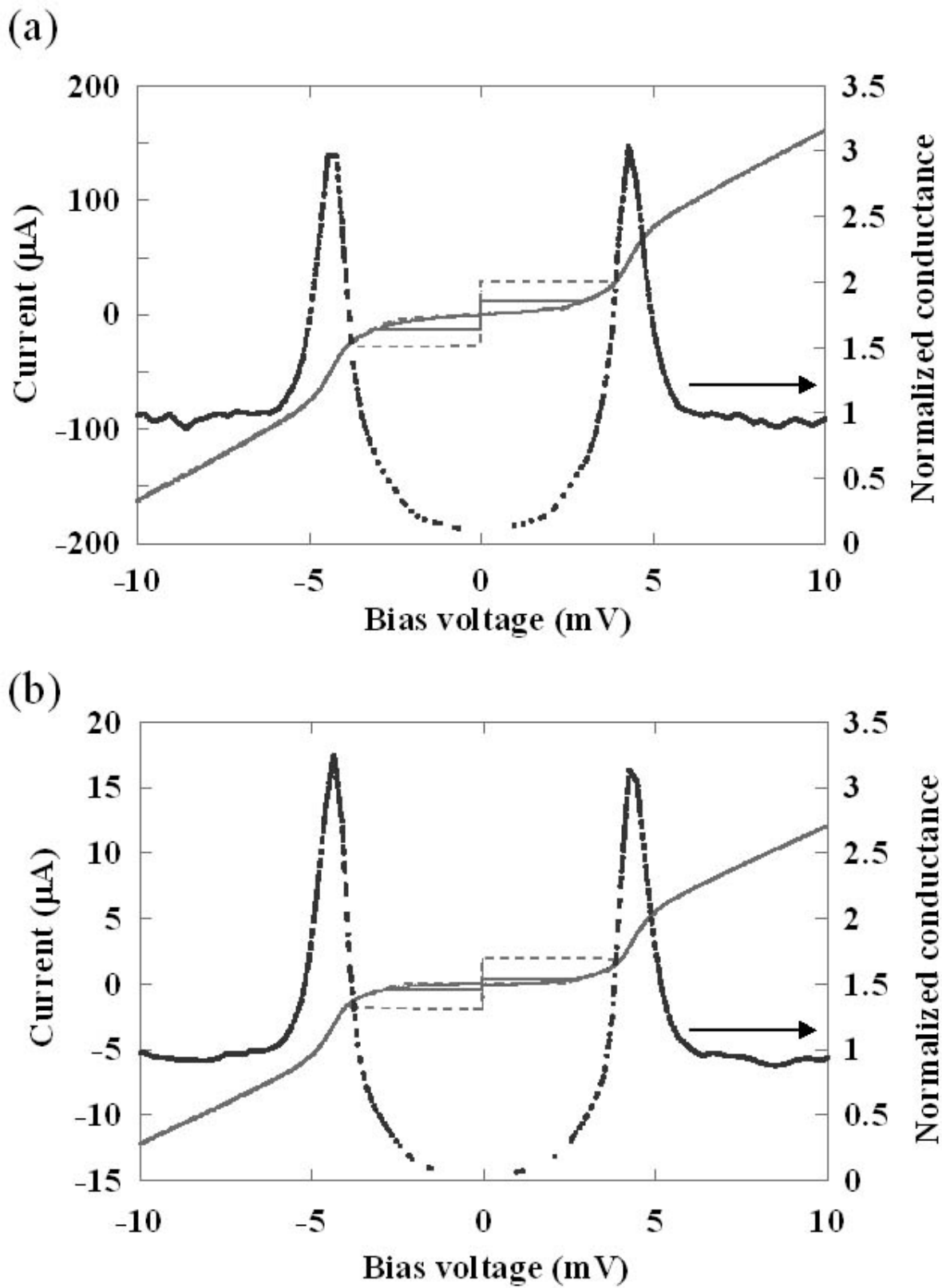

Fig. 1

K. Ueda, S. Saito, K. Semba, T. Makimoto, M. Naito

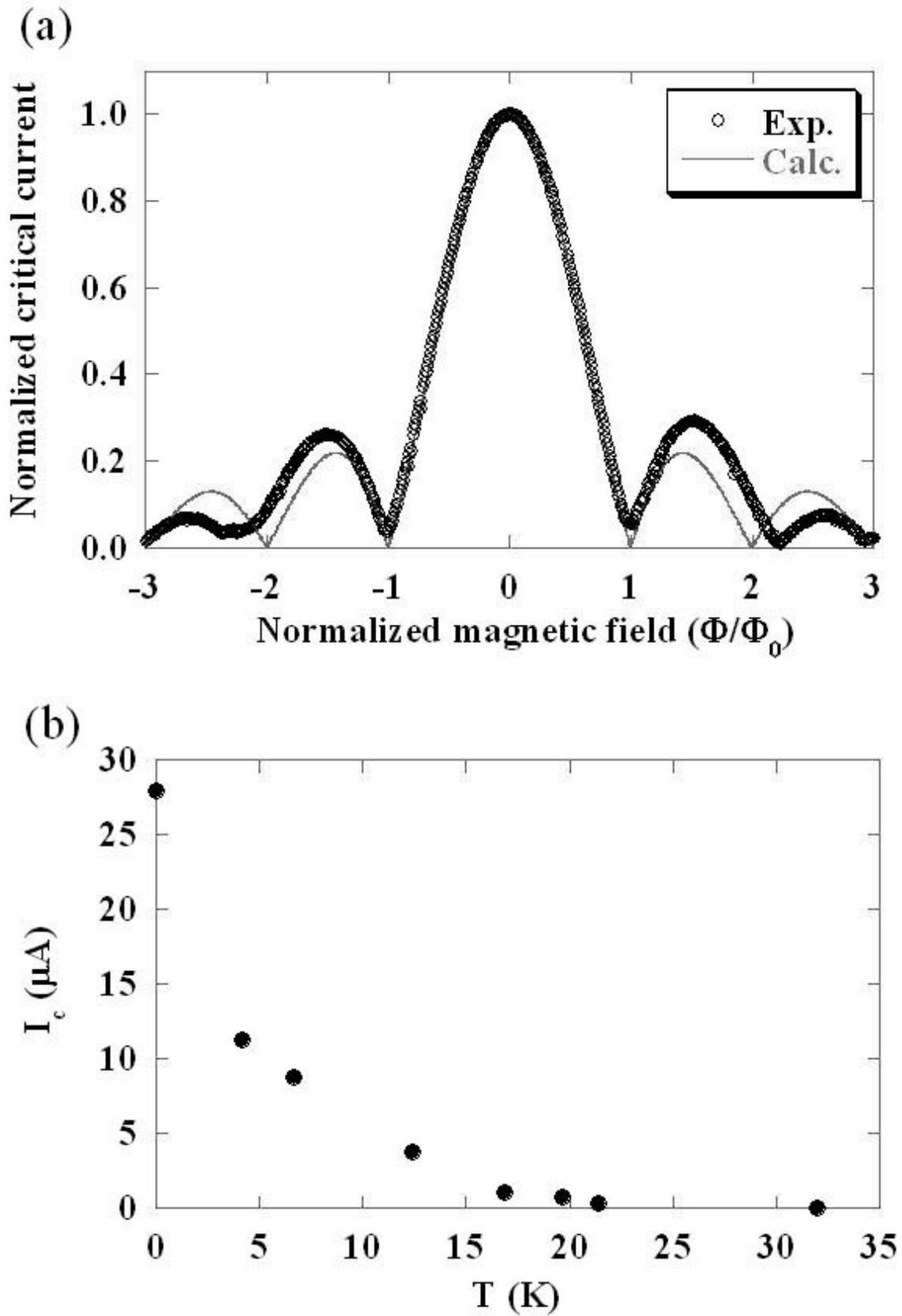

Fig. 2

K. Ueda, S. Saito, K. Semba, T. Makimoto, M. Naito

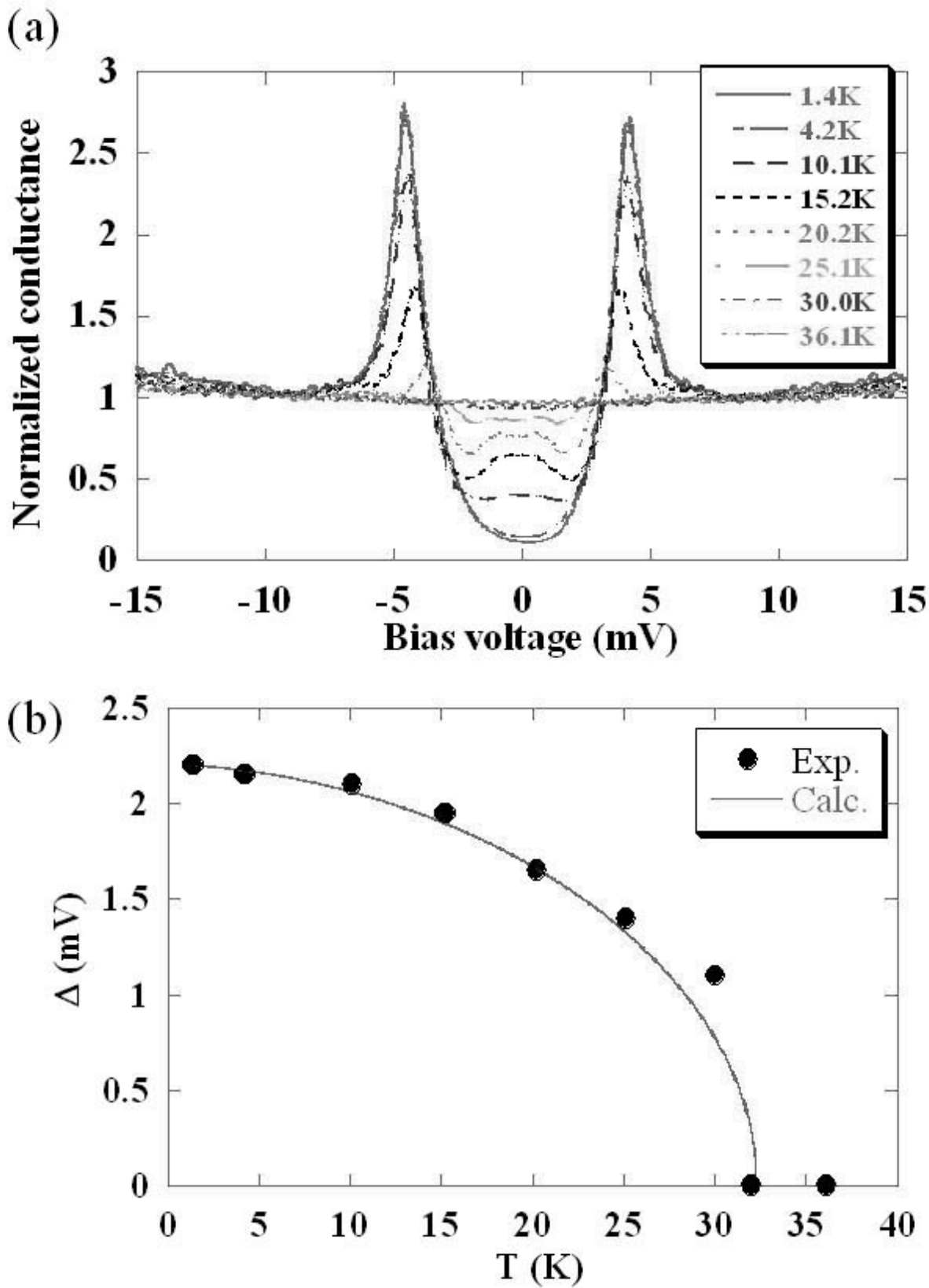

Fig. 3